\documentstyle[psfig]{l-aa}

\begin{document}

\thesaurus {06 (08.02.5)}

\title{Symbiotic stars on Asiago archive 
       plates\thanks{Table~1 is only available in electronic form
       at the CDS via anonymous ftp to cdsarc.u-strasbg.fr 
       (130.79.128.5) or via http://cdsweb.u-strasbg.fr/cgi-bin/qcat?J/A+A/}}

\author{
       Ulisse Munari\inst{1}
\and   Rajka Jurdana-\v{S}epi\'c\inst{2}
\and   Dina Moro\inst{1}
       }
\offprints{U.Munari}

\institute {
Osservatorio Astronomico di Padova, Sede di Asiago, 
I-36012 Asiago (VI), Italy
\and
Physics Department, University of Rijeka, Omladinska 14, HR-51000 Rijeka,
Croatia
}
\date{Received date..............; accepted date................}

\maketitle

\begin{abstract}
The rich plate archive of the Asiago observatory has been searched for
plates containing the symbiotic stars AS~323, Ap~3-1, CM~Aql, V1413~Aql (=
AS~338), V443~Her, V627~Cas (= AS 501) and V919~Sgr. The program objects
have been found on 602 plates, where their brightness has been estimated
against the {\sl UBV(RI)$_{\rm C}$} photometric sequences calibrated by
Henden and Munari (2000). 

AS~323 is probably eclipsing, with a preliminary P=197.6 day period. If
confirmed, it would be the shortest orbital period known among symbiotic
stars. CM~Aql does not seem to undergo a series of outbursts, its lightcurve
being instead modulated by a large amplitude sinusoidal variation with a
P$\sim$1058 day period. V627~Cas presents a secular trend in agreement with
the possible post-AGB nature of its cool component.

\end{abstract}
\maketitle

\section{Introduction}

The time scale of variability for symbiotic stars is quite long: the orbital
periods range from $\sim$1 year up to several decades while rise and decay
from an outburst may take anything from a few years to more than a century
(cf. Kenyon 1986).

Such long time scales tend to discourage stand-alone photometric campaigns
from a single Observatory, which could pay dividends only after ten or more
years. Most of the current photometric investigations of symbiotic stars
therefore try to assemble as much as possible data from the widest set of
current and archival sources. Template examples are the reconstruction of
the 1890-1996 lightcurve of YY~Her by Munari et al. (1997) and the 1885-1988
lightcurve of CH~Cyg by Mikolajewski et al. (1990). Both required a huge
effort in locating and measuring historical material in plate archives
around the world.

Henden and Munari (2000,2001) have so far provided accurate and extended
{\sl UBV(RI)$_{\rm C}$} photometric comparison sequences around 40 symbiotic
stars, intended to assist both present time photometry as well as measurement
of photographic plates from historical archives. They should stimulate small
observatories and/or occasional observers to obtain new data as well as to
encourage those with access to old plate archives to search for valuable
historical data. Assembling such data (obtained at various Observatories
against the same comparison sequences to minimize systematic errors) will
result in a much better understanding of the photometric evolution and
therefore the physical nature of this intriguing class of interacting
binaries.

In this paper we present the results of digging the Asiago plate archive for
seven symbiotic stars: AS~323, Ap~3-1, CM~Aql, V1413~Aql (= AS~338),
V443~Her, V627~Cas (= AS 501) and V919~Sgr.

\section{Data acquisition}

\begin{table*}
\caption[]{The {\sl UBV(RI)$_{\rm C}$} magnitudes of the program stars
estimated on the Asiago archive plates. The date is given in the
year/month/day format, the heliocentric JD is $HJD = JD_\odot -2400000$ and
the magnitude is estimated in steps of 0.05 mag.}
\centerline{\psfig{file=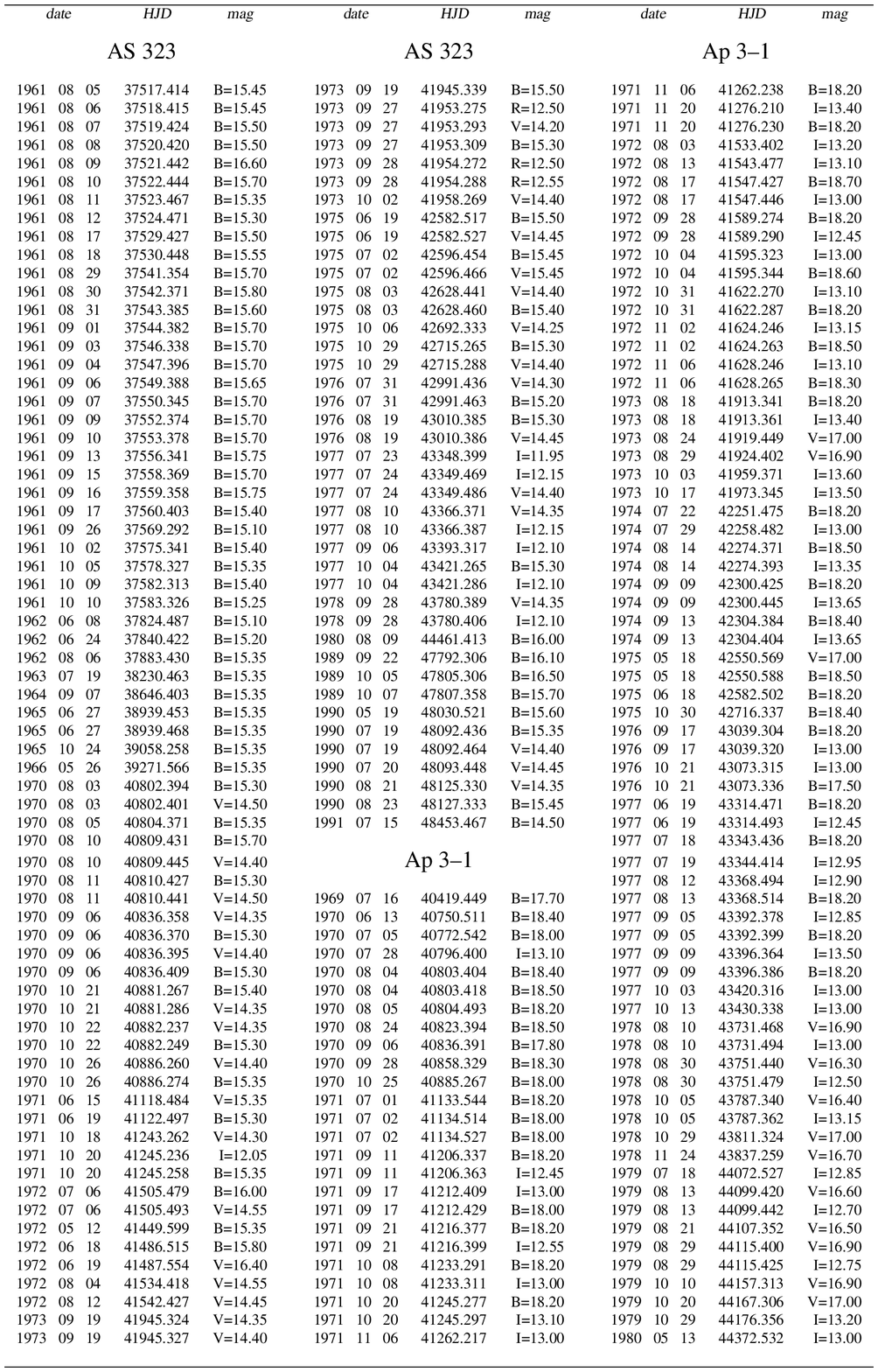,height=23.8cm}}
\end{table*}

\setcounter{table}{1}
\begin{table*}
\caption[]{({\sl continues})}
\centerline{\psfig{file=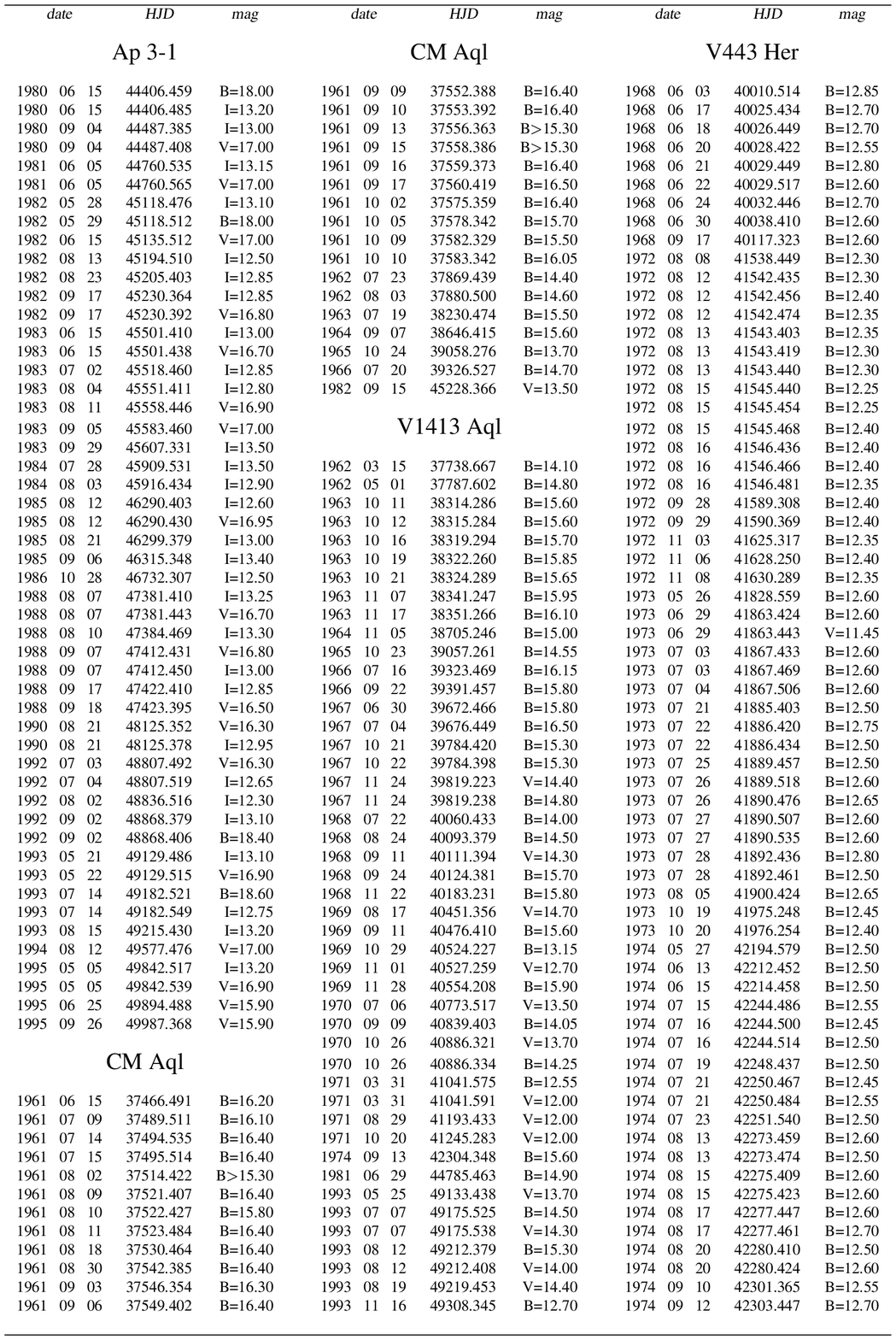,height=23.8cm}}
\end{table*}

\setcounter{table}{1}
\begin{table*}
\caption[]{({\sl continues})}
\centerline{\psfig{file=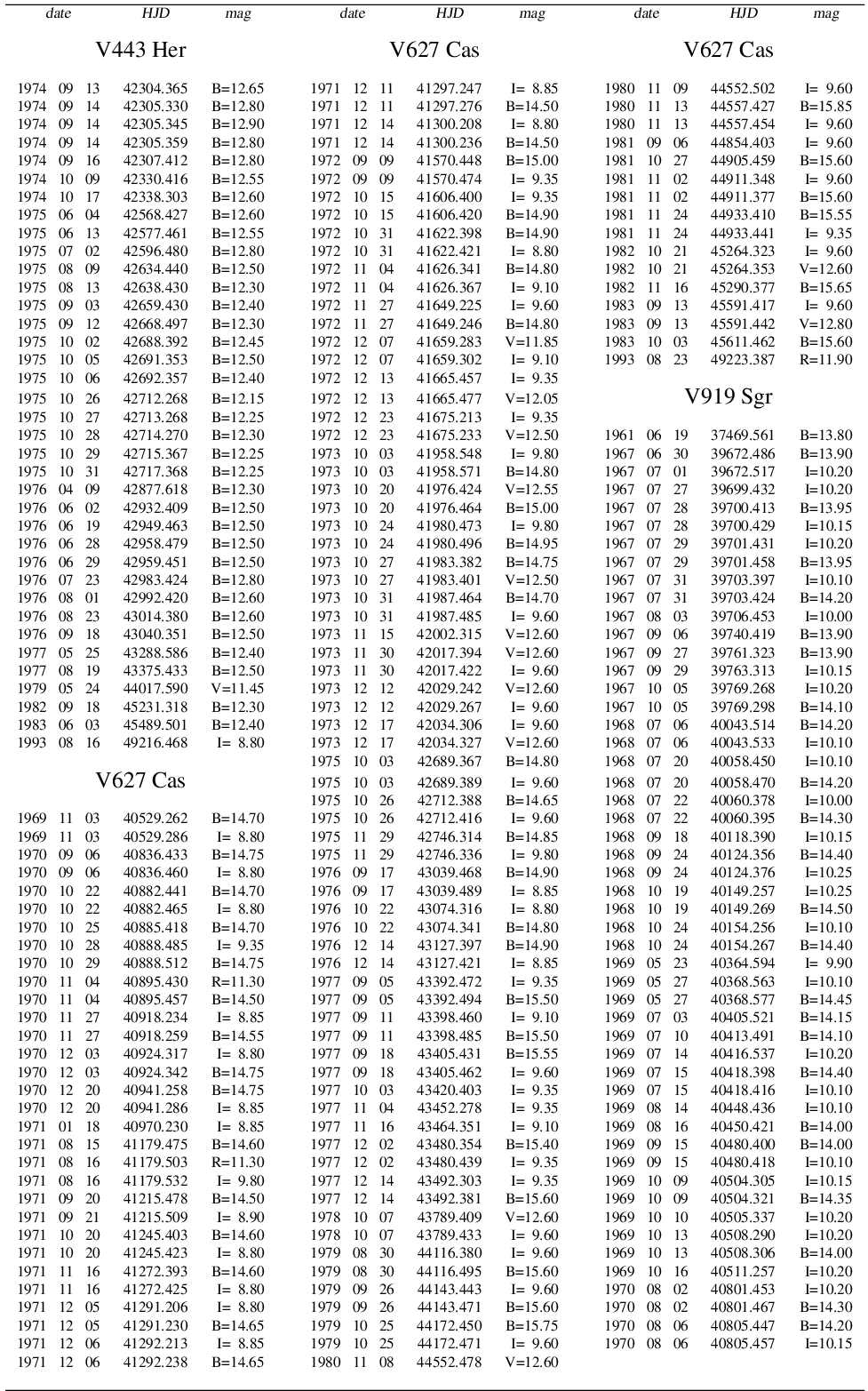,height=23.8cm}}
\end{table*}

Two Schmidt telescopes were operated at Asiago observatory. The smaller one
(40/50 cm, 100 cm focal length) collected 20417 plates from 1958 to 1992,
and the larger one (67/92 cm, 208 cm focal length) 18811 plates from 1965 to
1998. The Asiago Schmidt plate collection thus span 40 years. The majority
of the plates match the $B$ band, but the $U$, $V$, $R_C$ and $I_C$ bands
are well represented too.

The plates are typically filed in the archive logs with the coordinates of
the object to which they were aimed, that generally does not lay in the
plate center (which is instead usually the case for the guiding star).

Therefore, for a given program star, we initially selected from the archive
logs the plates to inspect as if they were covering a 2$\times$2 wider area. 
A subsequent visual inspection of all the selected plates separated those
actually containing the program star (602 plates) from the others.

We then proceeded to estimate at an high quality binocular microscope the
magnitude of the program star against the {\sl UBV(RI)$_{\rm C}$} comparison
sequences calibrated by Henden and Munari (2000). These comparison sequences
proved to work perfectly, covering the range of variability of the program
stars and with both the comparison stars and the variable visible at the
same time in the eyepiece field of the microscope.  The exception has been
V627~Cas, which was brighter than the comparison sequence in some of the $R$
and $I$ plates. We then searched outside the field explored by Henden and
Munari (2000) for bright stars that have been found constant in brightness
by Hipparcos/Tycho. We converted their Tycho $B_T$, $V_T$ magnitudes into
Johnson's standard $B$ and $V$ values, and using the transformations of
Caldwell et al. (1993) we eventually derived their $R_{\rm C}$ and $I_{\rm
C}$ magnitudes.  These transformation relations between colors in the
$UBV(RI)_{\rm C}$ system gives accurate results {\sl provided} that the
stars belong to the solar neighborhood population, the reddening is not
large and the luminosity class is roughly known. We have assumed all the
selected Tycho objects to be nearby main sequence stars. Thus, the $R_{\rm
C}$ and $I_{\rm C}$ so derived may be considered only as guidelines useful for
estimating photographic plates. The two stars we used to extend the $R_{\rm
C}$ and $I_{\rm C}$ comparison sequences around V627~Cas are TYC~3997~2203~1
($R_{\rm C}=9.11$ and $I_{\rm C}=9.10$) and TYC~3997~1868~1 ($R_{\rm
C}=8.71$ and $I_{\rm C}=8.67$).

The data are presented in Table~1. The date (year/ month/ day/ format), the
heliocentric JD and the estimated magnitude (in steps of 0.05 mag) are
given. Further details (including plate number, exposure time, emulsion and
filter types, etc.) are available via http://ulisse.pd.astro.it/symbio\_pg/.

\section{Notes on individual objects}

Brief notes follow to comment upon the photometric behavior displayed
by the program stars.

{\sl Ap~3-1} varied by $\bigtriangleup I \sim \bigtriangleup 
B \sim 1$ mag over the 26 years covered by the Asiago plates in Table~1, 
but without following any obvious periodic pattern or monotonic trend.

{\sl V1413~Aql}. The 1962-1981 sub-set of plates were already analyzed
by Munari (1992) against a different comparison sequence. The Table~1 data
confirm the pre-outburst lightcurve modulated by a reflection effect
following the 434.1 day orbital period of the post-outburst eclipses.

{\sl V443~Her}. 96 of the 100 $B$ band data in Table~1 cover the 
period 1968-1977. They confirm in periodicity (594 day) and amplitude
($\bigtriangleup B \sim 0.4$ mag) the later 1979-1993 $B$ lightcurve of
Kolotilov et al. (1995). Unavoidable small differences in pass-band
profiles between the photoelectric and photographic realization of the $B$
band can contribute to the slight difference in mean brightness
($B=$12.43 for Kolotilov et al. photoelectric photometry, $B=$12.53 for
Table~1 data).

{\sl V919~Sgr}. Our 1961-1970 data are affected only by a small
amplitude variability: $\bigtriangleup I \sim$0.3 and $\bigtriangleup 
B \sim$0.5 mag.

\subsection{AS~323}

Originally classified as a planetary nebula (= K~4--7 =
PK~26~$-2^{\circ}2$), the symbiotic nature of AS~323 has been discovered by
Sabbadin (1986) and Acker et al. (1988). The spectrum resembles a proto-type
symbiotic star, with well developed TiO bands in the red, veiling by the
circumstellar nebula in the blue and a high ionization emission line
spectrum (He~II 4686 \AA\ and 6825 \AA\ Raman scattering of O~VI are both
prominent), with weak or absent nebular lines. Mikolajewska et al. (1997)
estimated a M3 spectral type for the cool giant and $T_{eff}\geq$100\, 000~K
and $L=$1200 $L_\odot$ for the hot companion. Munari et al. (2001) report
$B=15.18$, $B-V=+0.99$, $U-B=-0.42$, $V-R_C=+1.11$ and $R_C-I_C=+1.39$ for
observations obtained in 1999.

\begin{figure}[!t]
\centerline{\psfig{file=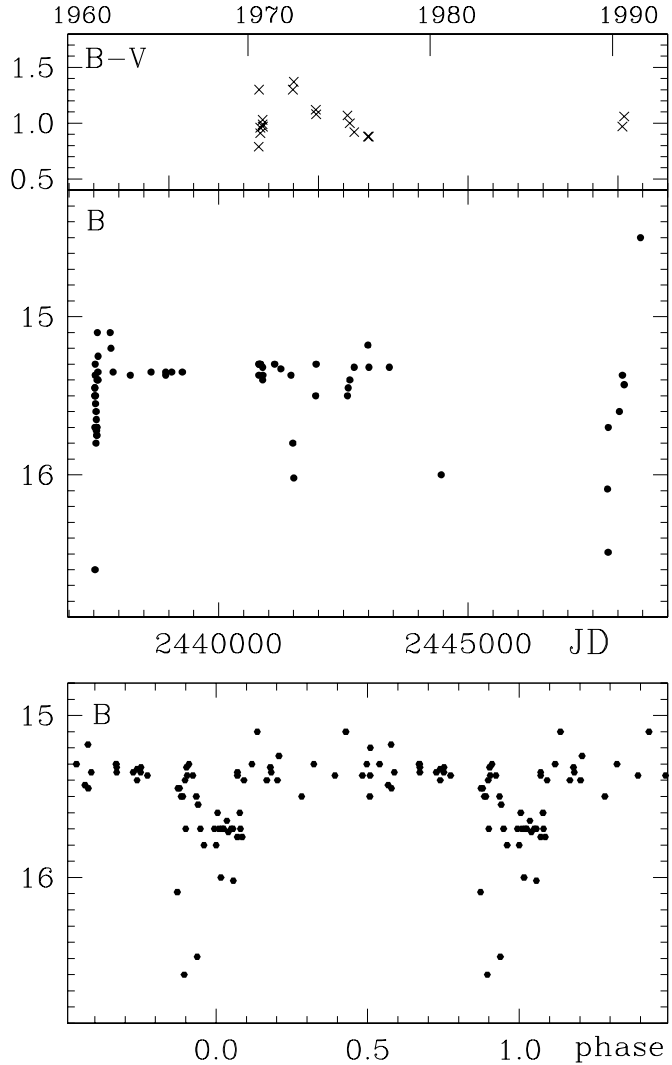,width=8.8cm}}
\caption[]{$B$ and {\sl B--V} lightcurves of AS~323 ({\sl upper panels}).
The $B$ data folded onto a P=197.6 day period ({\sl lower panel}).}
\end{figure}

The AS~323 data from Table~1 are plotted in Figure~1. The 1960-1990
lightcurve is characterized by a flat quiescence level at $B\sim$15.3, close
to the 1999 value. A sudden brightening occurred at the very end of the
observational period: its detection is solid given the careful check of the
AS~323 image on the plate that excluded local defects.

The most interesting aspect of the AS~323 lightcurve is however the series
of drops below the quiescent level. A search for periodicities has revealed
several possible periods, the stronger one being P=197.6 days. The $B$ data
are phase-plotted against it in Figure~1, showing a lightcurve closely
resembling a deep eclipsing binary ($\bigtriangleup B \sim 1.5$ mag). More
data are however necessary to firmly establish the periodicity, refine the
period and confirm the suspected eclipsing nature. Hopefully,
similar programs could locate in other archives more plates containing
AS~323 and solve the issue.

If the P=197.6 days should be confirmed as the orbital period of AS~323, it
would be the shortest known among symbiotic stars, with the closest cases
being TX CVn (199 days), T~CrB (228 days) and BD-21.3873 (282 days; cf.
Belczy\'{n}ski et al. 2000). The M3 giant in AS~323 would then quite
probably fill its Roche lobe and show the characteristic {\sl ellipsoidal }
distortion of its lightcurve.

\begin{figure}[!t]
\centerline{\psfig{file=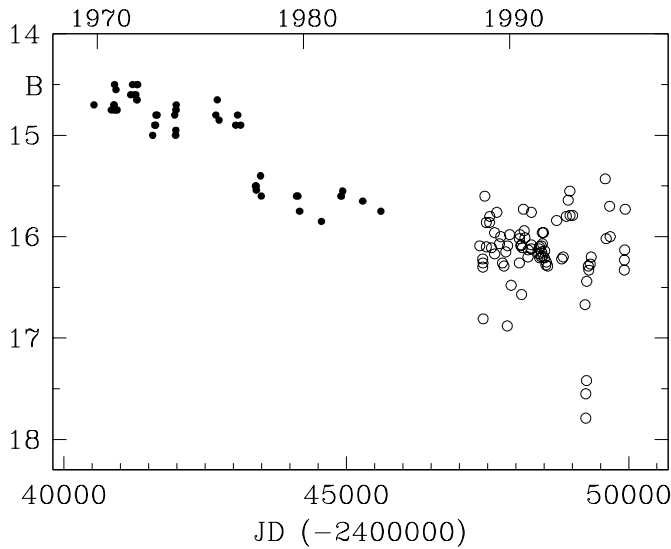,width=8.8cm}}
\caption[]{$B$ lightcurve of V627~Cas (=AS~501). {\sl Dots}: data from Table~1.
{\sl Open circles}: photoelectric photometry by Kolotilov et al. (1996).}
\end{figure}
\begin{figure}[!t]
\centerline{\psfig{file=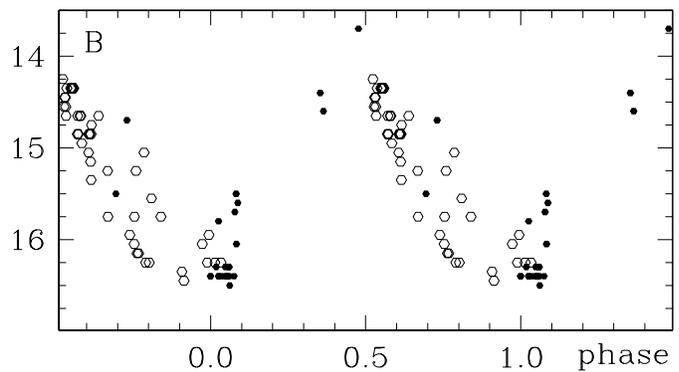,width=8.8cm}}
\caption[]{The $B$ data of CM~Aql folded onto a P=1058 day period. 
{\sl Dots}: data from Table~1.
{\sl Open circles}: data from Harwood (1925).}
\end{figure}

\subsection{V627~Cas}

According to Kolotilov et al. (1996), V627~Cas (=AS~501) is an unusual type
of symbiotic star because it could harbor a post-AGB cool giant. During the
post-AGB phase a star is supposed to evolve very rapidly. The secular
decrease in brightness evident in Figure~2 could be then ascribed to global
modifications of the cool giant that dominates the emission of V627~Cas in
the $B$ band.

Kolotilov et al. (1996) also reported on a small amplitude pulsation of the
cool giant with a period of P=466 days. This periodicity (or any other else)
does not seem to be present in the $B$ or $I$ band data of Table~1, which
cover an earlier time interval. The fact that the cool giant may have
started to pulsate on such short time scales may again argue in favor of the
Kolotilov et al. (1996) scenario of a post-AGB, rapidly evolving star.

\subsection{CM Aql}

According to the extensive literature search by Kenyon (1983), CM~Aql has
varied in the past between $16.4 \geq B \geq 13.2$, with outbursts recorded
in 1914, 1925 (when it was discovered), 1934 and 1950.

Our data in Table~1 extend over $16.5 \geq B \geq 13.7$, the same as in
older records, without evidence for separated quiescence and outburst phases
and favoring instead a continuous variability. It has also to be noticed
that at the time of the 1925 ``outburst" HeII~4686 was in strong emission
(Harwood 1925), contrary to the typical behavior of symbiotic stars.

We argue that the variability so far observed in CM~Aql is not modulated by
outbursts, but it is instead periodic in nature. In Figure~3 the
data of Harwood (1925) and those of Table~1 are phase plotted according to a
period of P=1058 days. The sinusoidal shape would suggest a {\sl reflection
effect} interpretation. A $\bigtriangleup B \sim$2 mag amplitude would
however be unusually large for a symbiotic star. Clearly, further data from other
plate archives are necessary to firmly address the period, the nature of
the sinusoidal-like variability and the absence of outbursts.

\end{document}